\documentclass[12pt]{article}
\usepackage{amssymb}
\usepackage{amsfonts}

\input{tcilatex}
\begin{document}

\begin{center}
\bigskip

\textbf{Hyperboloid preservation }

\textbf{implies the Lorentz and Poincar\'{e} groups without dilations}
\end{center}

\bigskip

\begin{center}
Stephan Foldes

Tampere University of Technology

PL 553, 33101 Tampere, Finland

sf@tut.fi

September 2010

\bigskip

\textbf{Abstract}
\end{center}

\textit{An analogue of the Alexandrov-Zeeman theorem, based on hyperboloid
preservation, as opposed to light cone preservation, is provided. This
characterizes exactly the Poincar\'{e} group, as opposed to the Poincar\'{e}
group extended by dilations. The hyperbolic analogue, as opposed to the
cone-based Alexandrov-Zeeman theorem, is also valid in the case of a single
space dimension. An orthochronous version holds as well, based on the
preservation of forward hyperboloid shells.}

\textit{\bigskip }


\bigskip

\bigskip

\textbf{1 Introduction and statement of Theorem 1}

\bigskip

We propose to show that the Alexandrov-Zeeman (A-Z) theorem admits a tighter
analogue, where the role played by light cones is taken by hyperboloids
defined in $(n+1)$-dimensional space-time $%
\mathbb{R}
\times 
\mathbb{R}
^{n}$ by%
\[
\left\{ (t,\mathbf{x)}\in 
\mathbb{R}
\times 
\mathbb{R}
^{n}:(t-t_{0})^{2}-(\mathbf{x-x}_{0}\mathbf{)}\cdot (\mathbf{x-x}_{0}\mathbf{%
)}=1\right\} 
\]%
For $n=3$, if the center $(t_{0},\mathbf{x}_{0})$ of the hyperboloid is the
origin $0\mathbf{0}=0000$, then the upper sheet of the hyperboloid
(characterized by $t>0$) represents the set of possible velocity
four-vectors with respect to proper time, or the set of all forward unit
tangent vectors of possible world-lines of particles with positive mass.
Alternatively, the upper sheet of the hyperboloid can be viewed as the mass
shell consisting of the possible momentum four-vectors of unit mass
particles.

\bigskip

The A-Z theorem can be formulated in several equivalent ways. The
formulation we give in this section is based on light cones, as in [A-CJM],
rather than on the causality relation as in ]Z], but it refers to the doubly
infinite (past and future) cone, as in Latzer's version [L]. Alexandrov's
and Zeeman's earlier versions correspond to the orthochronous formulation
given in Section 4. (See e.g. Goldblatt [G] for an overview.)

\bigskip

A \textit{Lorentz transformation }of $(n+1)$-dimensional space-time $%
\mathbb{R}
\times 
\mathbb{R}
^{n}$, $n\geq 1$ is any non-singular linear transformation of $%
\mathbb{R}
\times 
\mathbb{R}
^{n}$\ mapping each vector $(t,\mathbf{x)}=(t,x_{1},...,x_{n})$ to a vector $%
(s,\mathbf{y})$ of the same Minkowski norm, \textit{i.e.} such that $t^{2}-%
\mathbf{x}^{2}=t^{2}-\mathbf{x}\cdot \mathbf{x}%
=t_{1}^{2}-(x^{2}+...+x_{n}^{2})$ equals $s^{2}-\mathbf{y}^{2}.$ Lorentz
transformations constitute the \textit{Lorentz group} (sometimes called the
extended Lorentz group as improper transformations and time reversals are
not excluded). A \textit{translation }is a transformation of $%
\mathbb{R}
\times 
\mathbb{R}
^{n}$ of the form $\mathbf{v}\longmapsto \mathbf{v+b}$, where $\mathbf{b\in }
$ $%
\mathbb{R}
\times 
\mathbb{R}
^{n}$ is some fixed vector. A \textit{Poincar\'{e} transformation} is a
Lorentz transformation followed by a translation, and these transformations
constitute the \textit{Poincar\'{e} group. }Finally, a \textit{dilation} is
a transformation of $%
\mathbb{R}
\times 
\mathbb{R}
^{n}$, necessarily linear and non-singular, of the form $\mathbf{%
v\longmapsto }a\mathbf{v}$, where $a$ is some fixed positive real number.

\bigskip

\textbf{A-Z Theorem}\textit{\ }(Alexandrov [A1, A2, A-CJM], Zeeman [Z],
Latzer [L]) \ \textit{With respect to the light cone}%
\[
C=\left\{ (t,\mathbf{x)}\in 
\mathbb{R}
\times 
\mathbb{R}
^{n}:t^{2}-\mathbf{x}^{2}=0\right\} 
\]%
\textit{in} $(n+1)$\textit{-dimensional space-time}, \textit{let }$G$ 
\textit{be the group of bijective transformations }$f$ \textit{of }$%
\mathbb{R}
\times 
\mathbb{R}
^{n}$ \textit{satisfying}%
\[
f[C+\mathbf{v]=}C\mathbf{+}f(\mathbf{v)} 
\]%
\textit{for all }$\mathbf{v\in }$ $%
\mathbb{R}
\times 
\mathbb{R}
^{n}$. \textit{If }$n\geq 2$\textit{, then }$G$\textit{\ is generated by the
Poincar\'{e} group and the group of dilations.}

\bigskip

The present paper is mainly devoted to deriving the following analogue:%
\[
\]

\bigskip

\textbf{Theorem 1 }\textit{Let }$n\geq 1.$\textit{With respect to the
hyperboloid}%
\[
H=\left\{ (t,\mathbf{x)}\in 
\mathbb{R}
\times 
\mathbb{R}
^{n}:t^{2}-\mathbf{x}^{2}=1\right\} 
\]%
\textit{in} $(n+1)$\textit{-dimensional space-time, let }$G$ \textit{be the
group of bijective transformations }$f$ \textit{of }$%
\mathbb{R}
\times 
\mathbb{R}
^{n}$ \textit{satisfying}%
\[
f[H+\mathbf{v]=}H\mathbf{+}f(\mathbf{v)} 
\]%
\textit{for all }$\mathbf{v\in }$ $%
\mathbb{R}
\times 
\mathbb{R}
^{n}$. \textit{Then }$G$\textit{\ is exactly the Poincar\'{e} group.}

\bigskip

In a slightly different formulation, the A-Z theorem says that the stabilizer%
\textit{\ }$G_{0}$ of the origin in the group $G$ (defined with reference to
light cone preservation) is generated by the Lorentz group and the group of
dilations. Theorem 1 says that the stabilizer $G_{0}$ (where $G$ is now
defined with reference to hyperboloid preservation) is precisely the Lorentz
group. For recent references in connection with the Lorentz group and
physical applications see e.g. [HL1, HL2, U, M, F].

\bigskip

Both the A-Z theorem and Theorem 1 will be shown to be consequences of Lemma
1 below (Sections 2 and 3). In Section 4 we show that the orthochronous
version of Theorem 1, based on forward hyperboloid shells, also holds and
characterizes the orthochronous Lorentz and Poincar\'{e} groups (Theorem 2
in Section 4).

\bigskip

The theorems will follow from a sequence of propositions of combinatorial
nature about hyperboloids and the three types of lines in Minkowski
space-time. Each one of these propositions can be verified with standard
techniques of linear and convex geometry, using also appropriate Lorentz or
Poincar\'{e} transformations. The detail of arguments will be indicated for
some of the propositions. As restricting the discussion to $3+1$ dimensions
offers no advantage in the present context, we carry out the argument for
arbitrary dimension in full generality.

\bigskip

\bigskip

\textbf{2 Collineations}\bigskip

Recall that in the vector space $%
\mathbb{R}
\times 
\mathbb{R}
^{n},$ a non-zero vector $(t,\mathbf{x)}$ is classified as \textit{%
space-like, light-like} or \textit{time-like} according to whether $t^{2}-%
\mathbf{x}^{2}$ is negative, zero or positive. A line in $%
\mathbb{R}
\times 
\mathbb{R}
^{n}$, being a translate of a 1-dimensional subspace $V,$ is classified as 
\textit{space-like, light-like} or \textit{time-like} according to whether
the non-zero vectors in $V$ are space-like, light-like or time-like. (These
are also called \textit{separation lines, optical lines} and \textit{inertia
lines}, respectively.) Two distinct points (vectors) $\mathbf{v,w\in }$ $%
\mathbb{R}
\times 
\mathbb{R}
^{n}$ are in \textit{space-like, light-like} or \textit{time-like relative
position} according to whether the line $\left\{ a\mathbf{v+}(1-a)\mathbf{w:}%
\text{ }a\mathbf{\in 
\mathbb{R}
}\right\} $ through them is space-like, light-like or time-like.

\bigskip

A plane in $%
\mathbb{R}
\times 
\mathbb{R}
^{n},$\textit{\ i.e. }a translate of a 2-dimensional subspace, is a\textit{\
Lorentz plane }if it contains two non-parallel light-like lines.
Characterized alternatively, these are the planes that contain lines of all
the three kinds (space-like, light-like and time-like). See Goldblatt [G]
for a discussion of the different types of planes in space-time.

\bigskip

We shall make use of the facts that

\bigskip

\textit{(a)} in a Lorentz plane $P$ there are exactly two light-like lines $%
L $ and $L^{\prime }$ through every point $\mathbf{v,}$

\bigskip

\textit{(b)} the Lorentz plane $P$ through any two distinct intersecting
light-like lines $L$ and $L^{\prime }$ consists of the the union of all
space-like lines intersecting $(L\cup L^{\prime })$ in more than one point,
plus the intersection point of $L$ and $L^{\prime }.$

\bigskip

The following lemma can itself be regarded as a version of the A-Z theorem,
with stronger line preservation hypotheses but not depending on the
assumption of more than one space dimension.

\bigskip

\textbf{Lemma 1}\textit{\ \ Let }$n\geq 1,$ \textit{and let }$G$\textit{\ be
the group of all bijective transformations }$f$ \textit{of }$%
\mathbb{R}
\times 
\mathbb{R}
^{n}$\textit{\ such that for every set of points }$L\subseteq 
\mathbb{R}
\times 
\mathbb{R}
^{n}$

\textit{(i) } $L$ \textit{is a space-like line if and only if }$f[L]$\textit{%
\ is one,}

\textit{(ii) } $L$ \textit{is a light-like line if and only if }$f[L]$%
\textit{\ is one.}

\textit{Then }$G$\textit{\ is generated by the Poincar\'{e} group and the
group of dilations.}

\bigskip

\textbf{Proof} \ Clearly any $f\in G$ maps any pair $\left\{ \mathbf{v,w}%
\right\} $ of distinct points in space-like (respectively light-like)
relative position to a point-pair $\left\{ f(\mathbf{v}),f(\mathbf{w)}%
\right\} $ in space-like (respectively light-like) position, and thus it
must map point-pairs in time-like position to point-pairs in time-like
position.

We claim that actually for any $f\in G$ and subset $L\subseteq 
\mathbb{R}
\times 
\mathbb{R}
^{n}$, $L$ is a time-like line if and only if $f[L]$ is one. This, together
with conditions (i) and (ii) in the statement of the lemma, will imply that
any $f\in G$ is an affine transformation. To prove the claim, let $\mathbf{%
u,v,w\in }%
\mathbb{R}
\times 
\mathbb{R}
^{n}$ be three distinct points such that $\mathbf{u-v,v-w}$ and $\mathbf{u-w}
$ are space-like vectors, and let $P$ be the plane containing $\mathbf{u,v,w.%
}$ Then $P$ is a Lorentz plane. Let $L_{1},L_{2},L_{3}$ be parallel
light-like lines in $P$ through $\mathbf{u,v,w}$, respectively, and let $%
L_{1}^{\prime }\neq L_{1},L_{2}^{\prime }\neq L_{2},L_{3}^{\prime }\neq
L_{3} $ be the other light lines in $P$ through $\mathbf{u,v,w.}$ Let $%
K_{12},K_{13},K_{23}$ be the lines such that%
\begin{eqnarray*}
(L_{1}\cap L_{2}^{\prime })\cup (L_{1}^{\prime }\cap L_{2}) &\subseteq
&K_{12} \\
(L_{1}\cap L_{3}^{\prime })\cup (L_{1}^{\prime }\cap L_{3}) &\subseteq
&K_{13} \\
(L_{2}\cap L_{3}^{\prime })\cup (L_{2}^{\prime }\cap L_{3}) &\subseteq
&K_{23}
\end{eqnarray*}%
These lines are space-like. They are all parallel if $\mathbf{u,v,w}$ are
collinear, and no two of them are parallel otherwise. Any $f\in G$ must map
the Lorentz plane $P$\ to the Lorentz plane containing $f[L_{1}]\cup
f[L_{1}^{\prime }],$ and the points $f(\mathbf{u}),f(\mathbf{v}),f(\mathbf{w}%
)\in f[P]$ are collinear if and only if the lines $%
f[K_{12}],f[K_{13}],f[K_{23}]$ are all parallel, which proves the claim and
shows that every $f\in G$ is an affine transformation.

Take now any $f\in G.$ Let $\tau $ be the translation $\mathbf{v\mapsto v-}%
f(0\mathbf{0)}$. The composition $\tau f\in G$ is a linear transformation.
Let $\lambda $ be a Lorentz transformation sending $\tau f(0\mathbf{0)}$ to
a vector of the form $(a,\mathbf{0),}$ $a\neq 0$, and let $\delta $ be the
dilation $\mathbf{v\mapsto }a^{-1}\mathbf{v}$. Then $\delta \lambda \tau
f\in G$ is a linear transformation fixing $(1,\mathbf{0)}$ and it is easily
seen to keep the hyperplane $S=\left\{ (0,\mathbf{x}):\mathbf{x\in 
\mathbb{R}
}^{n}\right\} $ stable. It can be verified that each $(0,\mathbf{x})\in S$
must be sent by $\delta \lambda \tau f$ to a vector $\mathbf{(}0\mathbf{,%
\mathbf{y)}}$ satisfying $\left\Vert \mathbf{y}\right\Vert =\left\Vert 
\mathbf{x}\right\Vert $. Thus there is a Lorentz transformation $\rho $
keeping $S$ stable (space rotation) such that $\rho \delta \lambda \tau f$
is the identity, and $f=(\tau \lambda \delta \rho )^{-1}.$ \ \ \ \ \ $%
\square $

\bigskip

The above lemma implies the A-Z theorem as follows. \textit{Optical
hyperplanes }(hyperplanes containing space-like and light-like lines but no
time-like lines, also called \textit{Robb hyperplanes} in [G]) are
characterized in terms of light-like connection alone as sets of points in $%
\mathbf{%
\mathbb{R}
\times 
\mathbb{R}
}^{n}$ of the form%
\[
L\cup \left\{ \mathbf{v\in 
\mathbb{R}
\times 
\mathbb{R}
}^{n}:\text{ }\mathbf{v-w}\text{ \textit{is not light-like for any }}\mathbf{%
w}\in L\right\} 
\]%
where $L$ is any light-like line. Space-like lines are characterized as
minimal non-empty non-singleton intersections of Robb hyperplanes, if $n\geq
2.$ This fails for $n=1$, but for $n\geq 2,$ Lemma 1 can be applied to
obtain the A-Z theorem.

\bigskip

\bigskip

3 \textbf{Dilated hyperboloids and proof of Theorem 1}

\bigskip

First we shall characterize point pairs in light-like relative position in
terms of certain hyperboloids to which the two points belong (Lemma 2, its
Corollary, and Proposition 1 below). This will allow the application of the
A-Z theorem in the case of at least two space dimensions. In order to handle
the case of a single space dimension, we shall go back to Lemma 1 in the
previous section, but first we shall need to characterize also space-like
lines in terms of hyperboloids (Propositions 2-5). These latter propositions
will be stated and proved in full generality for any number $n\geq 1$ of
space dimensions.

\bigskip

For any fixed vector\ (space-time point) $\mathbf{v\in 
\mathbb{R}
\times 
\mathbb{R}
}^{n}$ and real number $r>0$ consider the hyperboloid%
\[
H(\mathbf{v},r)=\left\{ (t,\mathbf{x)}\in 
\mathbb{R}
\times 
\mathbb{R}
^{n}:t^{2}-\mathbf{x}^{2}=r^{2}\right\} 
\]%
In particular, $H(\mathbf{0},1)$ is the hyperboloid $H$ appearing in the
statement of Theorem 1. For any set of points $S\subseteq \mathbf{%
\mathbb{R}
\times 
\mathbb{R}
}^{n}$ and real number $m>0$, consider the dilation of $S$ by the factor $m,$%
\[
mS=\left\{ m\mathbf{u:u\in }S\right\} 
\]%
Note that for any positive real numbers $m$ and $r$ we have $mH(\mathbf{0}%
,r)=H(\mathbf{0},mr)$.

\bigskip

\textbf{Lemma 2}\textit{\ \ }\ \textit{Let} $n\geq 1$. \textit{For any} 
\textit{positive real number} $r$ \textit{we have}

(i) \ \ \ \ \ \ \ \ \ \ $2H(\mathbf{0},r)=\left\{ \mathbf{v:}\text{ }H(%
\mathbf{v},r)\cap H(\mathbf{0},r)\text{ is a singleton}\right\} $

(ii) \ \ \ \ \ \ \ \ \ $\ H(\mathbf{0},r)=\left\{ \mathbf{v:}\text{ }H(%
\mathbf{v},r)\cap 2H(\mathbf{0},r)\text{ \textit{is a singleton}}\right\} $

\bigskip

\textbf{Proof} \ (i) \ To show that for every $\mathbf{v\in }$ $2H(\mathbf{0}%
,r)$ the intersection $H(\mathbf{v},r)\cap H(\mathbf{0},r)$ is a singleton,
it is enough to see that this holds for $\mathbf{v=}(2r,\mathbf{0),}$
because the Lorentz group acts transitively on $2H(\mathbf{0},r)$, and any
Lorentz transformation mapping $\mathbf{u}$ to $\mathbf{v}$ maps $H(\mathbf{u%
},r)$ to $H(\mathbf{v},r)$ and $H(\mathbf{0},r)$ to itself. It can also be
seen easily that for all $t\geq 0,$ $t\neq 2r,$ the intersection $%
H((t,0...0),r)\cap H(\mathbf{0},r)$ is either infinite or empty, and this,
combined again with a transitivity consideration, shows that $H(\mathbf{v}%
,r)\cap H(\mathbf{0},r)$ is a singleton only if $\mathbf{v\in }$ $2H(\mathbf{%
0},r).$

(ii) This is proved similarly. $\square $

\bigskip

\textbf{Corollary }\ \textit{Let} $n\geq 1$\textit{\ and consider the
hyperboloid }$H=H(\mathbf{0,}1).$ \textit{If a bijective transformation }$f$%
\textit{\ of} $\mathbf{%
\mathbb{R}
\times 
\mathbb{R}
}^{n}$ \textit{satisfies }%
\[
f[H-\mathbf{v}]=H+f(\mathbf{v}) 
\]%
\textit{for all} $\mathbf{v\in 
\mathbb{R}
\times 
\mathbb{R}
}^{n},$ \textit{then it also satisfies }%
\[
f[2^{e}H-\mathbf{v}]=2^{e}H+f(\mathbf{v}) 
\]%
\textit{for all} $\mathbf{v\in 
\mathbb{R}
\times 
\mathbb{R}
}^{n}$ \textit{and all positive or negative integer exponents }$e\in 
\mathbb{Z}
$. $\ \square $

\bigskip

\textbf{Proposition 1}\textit{\ \ Two distinct points }$\mathbf{u,w\in 
\mathbb{R}
\times 
\mathbb{R}
}^{n},$ $n\geq 1,$ \textit{are in light-like relative position if and only
if for each integer }$e\in 
\mathbb{Z}
$ \textit{and }$\mathbf{v\in 
\mathbb{R}
\times 
\mathbb{R}
}^{n}$, \textit{at most one of }$\mathbf{u,w}$\textit{\ lies on the
hyperboloid }$H(\mathbf{v,}2^{e}).$

\bigskip

\textbf{Proof \ }First, no two distinct points in light-like relative
position lie on any hyperboloid $H(\mathbf{v},r)$.

Second, for every positive real $x$ there is a positive real $t$ such that
both $(t,-x,0,...,0\mathbf{)}$ and $(t,x,0,...,0\mathbf{)}$ lie on $H(%
\mathbf{0},1)$ \ By translation and Lorentz transformation, any two points
is space-like relative position lie on some $H(\mathbf{v},1)$

Third, let $t>0.$ Take any integer $e$ such that $2^{e}\leq t$. Then there
is a positive real $x$ such that both $(t,-x,0,...,0\mathbf{)}$ and $%
(t,x,0,...,0\mathbf{)}$ lie on $2^{e}H(\mathbf{0},1)=H(\mathbf{0},2^{e}).$
Again by translation and Lorentz transformation, any two points in time-like
relative position line on some $H(\mathbf{0},2^{e})$ for an appropriate $e.$ 
$\ \ \ \ \square $

\bigskip

It is now clear that a transformation $f$\ belonging to the hyperboloid
preserving group $G$ of Theorem 1 satisfies for all points $\mathbf{v}$ the
light cone preservation condition $f[C+\mathbf{v]=}C\mathbf{+}f(\mathbf{v)}$
appearing in the A-Z theorem. Also, hyperboloid preservation rules out
non-trivial dilations. This proves Theorem 1 for all $n\geq 2.$ In order to
include also the case $n=1$ we shall characterize, in terms of hyperboloids,
first point-pairs in space-like position, then space-like collinearity. An
argument along the second part of the proof of Proposition 1 shows the
following:

\bigskip

\textbf{Proposition 2 \ }\textit{Two distinct points }$\mathbf{u,w\in 
\mathbb{R}
\times 
\mathbb{R}
}^{n},$ $n\geq 1,$ \textit{are in space-like relative position if and only
if for all integers }$z\in 
\mathbb{Z}
$ \textit{and }$\mathbf{v\in 
\mathbb{R}
\times 
\mathbb{R}
}^{n},$\textit{\ there is a }$\mathbf{v\in 
\mathbb{R}
\times 
\mathbb{R}
}^{n},$ \textit{such that both }$\mathbf{u,w}$\textit{\ lie on the
hyperboloid }$H(\mathbf{v,}2^{e}).$ $\ \square $

\bigskip

Each hyperboloid $H(\mathbf{v},r)$ has two topologically connected
components, called \textit{shells.} A hyperboloid shell $C$\ is \textit{%
forward }or \textit{backward} according to whether the time projection $%
\left\{ t\in 
\mathbb{R}
:(t,\mathbf{x})\in C\right\} $ is bounded from below or from above. Two
shells have the \textit{same orientation} if both are forward or both are
backward, otherwise they have \textit{opposite orientation}. For each shell $%
C$ there is only one point $\mathbf{v}$ and only one positive constant $r$
such that $C$ is one of the two shells of opposite orientation making up the
hyperboloid $H(\mathbf{v},r)$: the constant $r$ and the point $\mathbf{v}$
are the \textit{radius} and the \textit{center} of the shell $C.$ If the
radius $r$ is an integer power of $2,$ \textit{i.e.} if $r=2^{e}$ for some $%
e\in 
\mathbb{Z}
$, then we shall call $C$ a \textit{standard shell}.

\bigskip

\textbf{Proposition 3 \ }\textit{Two distinct points }$\mathbf{u,v}\in 
\mathbf{%
\mathbb{R}
\times 
\mathbb{R}
}^{n},$ $n\geq 1,$ \textit{lying} \textit{on a hyperboloid }$H(\mathbf{v}%
,r), $ \textit{lie on the same connected component of this hyperboloid if
and only if they are in space-like relative position. \ }$\square $

\bigskip

\textbf{Proposition 4 \ }\textit{In }$\mathbf{%
\mathbb{R}
\times 
\mathbb{R}
}^{n},$ $n\geq 1,$\textit{\ let }$C$ \textit{be a hyperboloid shell with
center} $\mathbf{v}$\ \textit{and }$K$\textit{\ a shell with center} $%
\mathbf{w}$\textit{. Assume that the centers }$\mathbf{v}$\textit{\ and }$%
\mathbf{w}$\textit{\ are in space-like relative position. Then }$C\cap
K=\emptyset $ \textit{if and only if }$C$\textit{\ and }$K$ \textit{have
opposite orientation. \ }$\square $

\bigskip

A point $\mathbf{v\in 
\mathbb{R}
\times 
\mathbb{R}
}^{n}$ is \textit{between} \textit{points} $\mathbf{u,w}$ if $\mathbf{v=}$ $a%
\mathbf{u}+(1-a)\mathbf{w}$ for some $0\leq a\leq 1.$ This implies that the
three points are collinear. Conversely, of any three collinear points one is
always between the other two.

\bigskip

\textbf{Proposition 5 \ }\textit{Let }$\mathbf{u,v,w}$ \textit{be three
distinct points in\ }$\mathbf{%
\mathbb{R}
\times 
\mathbb{R}
}^{n},n\geq 1,$ \textit{each two of them being in space-like relative
position. Then }$\mathbf{v}$ is \textit{between} $\mathbf{u}$ \textit{and}$\ 
\mathbf{w}$ \textit{if and only if the following holds:}

\textit{Whenever }$\mathbf{u,w}$\textit{\ lie on some standard hyperboloid
shell\ }$C$\textit{\ and }$\mathbf{v}$ \textit{lies on some standard shell }$%
K,$ \textit{these respective shells }$C$\textit{\ and }$K$\textit{\ are of
the same orientation or they possess a common point.}

\bigskip

\textbf{Proof} \ Suppose $\mathbf{v}$ is between $\mathbf{u}$ and $\mathbf{w}
$ but there are disjoint hyperboloid shells $C$ and $K$ of opposite
orientation with $\mathbf{u,w}\in C$ and $\mathbf{v}\in K.$ By an
appropriate Poincar\'{e} transformation we can map $\mathbf{u,v,w}$ to
points $\mathbf{u}^{\prime },\mathbf{v}^{\prime },\mathbf{w}^{\prime }$ such
that%
\[
\mathbf{u}^{\prime }=(t,-x,0,...,0\mathbf{)}\text{ \ \ }\mathbf{w}^{\prime
}=(t,x,0,...,0\mathbf{)} 
\]%
for some positive real numbers $t$ and $x$. Then $\mathbf{v}^{\prime
}=(t,q,0,...,0\mathbf{)}$ with $-x<q<x.$ The Poincar\'{e} transformation
used maps $C$ and $K$ to disjoint hyperboloid shells $C^{\prime }$ and $%
K^{\prime }$ which are also of orientation opposite to each other. But $%
\mathbf{u}^{\prime }\mathbf{,w}^{\prime }\in C^{\prime }$ and $\mathbf{v}%
^{\prime }\in K^{\prime }$, and the shells $C^{\prime }$ and $K^{\prime }$
must have a common point in the plane $\left\{ (t,x_{1},0,...,0\mathbf{):}%
\text{ }t,x_{1}\in 
\mathbb{R}
\right\} ,$ a contradiction proving the "common orientation or common point"
condition.

Suppose that $\mathbf{v}$ is not between $\mathbf{u}$ and $\mathbf{w}$. By
an appropriate Poincar\'{e} transformation $f$ we can map $\mathbf{u,v,w}$
to points $\mathbf{u}^{\prime },\mathbf{v}^{\prime },\mathbf{w}^{\prime }$
of the form%
\[
\mathbf{u}^{\prime }=(t,-x,0,...,0\mathbf{)}\text{ \ \ \ \ \ \ }\mathbf{w}%
^{\prime }=(t,x,0,...,0\mathbf{)}\text{ \ \ \ \ \ }t,x>0\text{\ \ \ \ \ \ \ }%
\mathbf{v}^{\prime }=(x_{0},x_{1},...,x_{n}) 
\]%
and such that either%
\[
\text{\ \ \ \ \ \ \ \ \ \ }x_{0}=t\text{ \ \ \ \ \ \ \ \ \ }x_{1}\notin
\lbrack -x,x]\text{ \ \ \ \ \ \ \ }x_{2}=...=x_{n}=0 
\]%
or $x_{0}\neq t.$ Clearly in both cases there exist two disjoint standard
hyperboloid shells $C^{\prime }$\ and $K^{\prime }$ of opposite orientation
such that $\mathbf{u}^{\prime }\mathbf{,w}^{\prime }\in C^{\prime }$ and $%
\mathbf{v}^{\prime }\in K^{\prime }.$ Then the inverse image shells $%
C=f^{-1}[C^{\prime }]$ and $K=f^{-1}[K^{\prime }]$ containing $\mathbf{u,v}$
and$\ \mathbf{w}$, respectively,\ are also disjoint and of opposite
orientation. $\square $

\bigskip

Suppose that a transformation $f$ belongs to the hyperboloid preserving
group $G$ of Theorem 1, and that the number $n$ of space dimensions is any
positive integer, possibly equal to $1.$ By Propositions 2 and 3, and the
Corollary of Lemma 2, $f$ must map standard hyperboloid shells to standard
shells. Using Propositions 4 and 5 it can be concluded that $f$ maps
space-like lines to space-like lines. Theorem 1 now follows from Lemma 1.

\bigskip

\bigskip

\textbf{4 Time orientation}

\bigskip

A Lorentz transformation is \textit{orthochronous} (preserves time
orientation) if it maps the vector $(1,\mathbf{0)}$ to a vector $(t,\mathbf{%
x)}$ with $t>0$. These transformations constitute the \textit{orthochronous
Lorentz group}.\ Together with all translations this group generates the 
\textit{orthochronous Poincar\'{e} group}, consisting of \textit{%
orthochronous Poincar\'{e} transformations. }Clearly all of these preserve
forward light cones and forward hyperboloids. The following corresponds to
Alexandrov's and Zeemans's original formulations of A-Z:

\bigskip

\textbf{A-Z Theorem, Orthochronous}\textit{\ }\textbf{Formulation}\textit{\ }%
(Alexandrov[A1, A2, A-CJM], Zeeman [Z]) \ \textit{With respect to the
forward light cone}%
\[
C^{+}=\left\{ (t,\mathbf{x)}\in 
\mathbb{R}
\times 
\mathbb{R}
^{n}:\text{ }t^{2}-\mathbf{x}^{2}=0\text{, \ }t\geq 0\right\} 
\]%
\textit{in} $(n+1)$\textit{-dimensional space-time}, \textit{let }$G$ 
\textit{be the group of bijective transformations }$f$ \textit{of }$%
\mathbb{R}
\times 
\mathbb{R}
^{n}$ \textit{satisfying}%
\[
f[C^{+}+\mathbf{v]=}C^{+}\mathbf{+}f(\mathbf{v)} 
\]%
\textit{for all }$\mathbf{v\in }$ $%
\mathbb{R}
\times 
\mathbb{R}
^{n}$. \textit{If }$n\geq 2$\textit{, then }$G$\textit{\ is generated by the
orthochronous Poincar\'{e} group and the group of dilations.}

\bigskip

(The deduction from the version stated in Section 1 is easy, as a
transformation $f$ preserving forward light cones necessarily preserves
light-like lines and thus preserves light cones. Any such $f$ is then a
dilation followed by a Poincar\'{e} transformation. If the dilation factor
is $a,$ then the Lorentz transformation $\mathbf{v\mapsto }a^{-1}\mathbf{[}f%
\mathbf{(v)-}f(0\mathbf{,0)]}$ is obviously orthochronous.)

\bigskip

Considering hyperboloid shells instead of cones, we obtain the orthochronous
version of Theorem 1:

\bigskip

\textbf{Theorem 2 \ }\textit{Let }$n\geq 1.$\textit{With respect to the
forward hyperboloid shell}%
\[
H^{+}=\left\{ (t,\mathbf{x)}\in 
\mathbb{R}
\times 
\mathbb{R}
^{n}:\text{ }t^{2}-\mathbf{x}^{2}=1,\text{ \ }t>0\right\} 
\]%
\textit{in} $(n+1)$\textit{-dimensional space-time, let }$G$ \textit{be the
group of bijective transformations }$f$ \textit{of }$%
\mathbb{R}
\times 
\mathbb{R}
^{n}$ \textit{satisfying}%
\[
f[H^{+}+\mathbf{v]=}H^{+}\mathbf{+}f(\mathbf{v)} 
\]%
\textit{for all }$\mathbf{v\in }$ $%
\mathbb{R}
\times 
\mathbb{R}
^{n}$. \textit{Then }$G$\textit{\ is exactly the orthochronous Poincar\'{e}
group.}

\bigskip

\textbf{Proof }The backward shell $H^{-}=H(\mathbf{0,}1)\setminus H^{+}$ $=%
\mathbf{-}H^{+}$ also satisfies%
\[
f[H^{-}+\mathbf{v]=}H^{-}\mathbf{+}f(\mathbf{v)} 
\]%
and therefore%
\[
f[H(\mathbf{0},1)+\mathbf{v]=}H(\mathbf{0},1)\mathbf{+}f(\mathbf{v)} 
\]%
for all\textit{\ }$\mathbf{v\in }$ $%
\mathbb{R}
\times 
\mathbb{R}
^{n}$ and all $f\in G.$ By Theorem 1, $G$ is a subgroup of the Poincar\'{e}
group, and it obviously contains all orthochronous Poincar\'{e}
transformations. But it can only contain orthochronous transformations,
because for every $f\in G$ the Lorentz transformation $\mathbf{v\mapsto }f%
\mathbf{(v)-}f(0\mathbf{,0)}$ is orthochronous. $\square $

\bigskip

\bigskip

\textbf{5 \ Concluding remarks}

\bigskip

Theorems 1 and 2 essentially say that for space-time transformations, not
assumed a priori to be linear or continuous, preservation of hyperboloids
(respectively of forward hyperboloid shells) implies light cone preservation
(respectively forward light cone preservation, i.e. preservation of
causality). The discussion was carried out for $%
\mathbb{R}
\times 
\mathbb{R}
^{n}$ without restriction on $n$, and without emphasizing the particular
significance of the case $n=3$.\ Copies of all lower dimensional space-time
models are embedded in all higher dimensional models. The case $n=1$
exhibits some specifics which can be dealt with either by restricting
attention to $n\geq 2$ as in the A-Z theorem, or by going around this result
as we have done in Section 3 to establish Theorem 1 for all $n\geq 1.$ We
note that the specifics of the case $n=1$ allow the consideration of
superluminal $(1+1)$-dimensional frames of reference, when viewed in
themselves and not as embedded in $(2+1)$ or $(3+1)$ dimensional space-time
(see Parker [P]).

\bigskip

Theorem 1 is formulated with reference to the hyperboloid $H=H(\mathbf{0},1)$
of radius $1$ whose upper component constitutes the unit mass shell $H^{+}$
that is shown by Theorem 2 to be the basic geometric invariant of the
orthochronous Lorentz group. It is straightforward to see that these
theorems also hold if reformulated with reference to any of the hyperboloids 
$H=H(\mathbf{0},m)$, $m>0.$

\bigskip

\bigskip

\textbf{References}

\bigskip

[A1] A.D. Alexandrov, On Lorentz transformations, Sessions Math. Seminar,
Leningrad Section of the Mathematical Institute, 15 September 1949
(abstract, in Russian)

\bigskip

[AO] A.D. Alexandrov, V.V. Ovchinnikova, Note on the foundations of
relativity theory, Vestnik Leningrad Univ. 11 (1953) 95-100 (in Russian)

\bigskip

[A-CJM] A.D. Alexandrov, A contribution to chronogeometry, Canadian J. Math.
19 (1967) 1119-1128

\bigskip

[F] S. Foldes, The Lorentz group and its finite field analogs: local
isomorphism and approximation, J. Mathematical Physics 49 (9) (2008)
093512:1-10.

\bigskip

[G] R. Goldblatt, Orthogonality and Spacetime Geometry, Springer 1987

\bigskip

[HL1] \ H.-K. Hong, C.-S. Liu, Lorentz group on Minkowski spacetime for
construction of the two basic principles of plasticity, Int. J. Non-Linear
Mechanics 36 (2001) 679-686

\bigskip

[HL2] \ H.-K. Hong, C.-S. Liu, Some physical models with Minkowski spacetime
structure and Lorentz group symmetry, Int. J. Non-Linear Mechanics 36 (2001)
1075-1084

\bigskip

[L] R.W. Latzer, Non-directed light signals and the structure of time, 
\textit{in} Space, Time and Geometry, P. Suppes (ed.) D. Reidel 1973, pp.
321-365

\bigskip

[M] V. Moretti, The interplay of the polar decomposition theorem and the
Lorentz group, Lecture Notes, Seminario Interdisciplinare di Matematica
5-153 (2006) 18 pages, also ArXiv:math-ph/0211047v1 at www.arxiv.org

\bigskip

[P] L. Parker, Faster-than-light inertial frames and tachyons, Physical
Review 188 No. 5 (1969) 2287-2292

\bigskip

[U] H.K. Urbantke, Lorentz transformations from reflections: some
applications, Found. Phys. Lett. 16 (2003) 111-117

\bigskip

[Z] E.C. Zeeman, Causality implies the Lorentz group, J. Mathematical
Physics 5 (1964) 490-493

\end{document}